\documentstyle[twocolumn,psfig]{mn}

\newif\ifAMStwofonts
\AMStwofontstrue

\def\mpc{\rm{h^{-1}Mpc}}

\def\gsim{~\rlap{$>$}{\lower 1.0ex\hbox{$\sim$}}}

\def\xm{x_{_{\rm mond}}}

\def\ltsim{\lower.5ex\hbox{$\; \buildrel < \over \sim \;$}}
\def\gtsim{\lower.5ex\hbox{$\; \buildrel > \over \sim \;$}}
\def\ltsim{\lower.5ex\hbox{$\; \buildrel < \over \sim \;$}}
\def\gtsim{\lower.5ex\hbox{$\; \buildrel > \over \sim \;$}}

\def\vx{{\bf x}}

\def\vr{{\bf r}}

\def\vu{{\bf u}}
\def\vb{{\bf b}}

\def\pa{\partial}
 
\def\vg{{\bf g}}
\def\vh{{\bf h}}
\def\vnabla{{\bf \nabla}}

\def\dd{\,{\rm d}}

\begin{document}
\title[MOND of LSS]{Modified Newtonian Dynamics of Large Scale Structure}

\author[Nusser ] { Adi Nusser  \\
  Physics Department- Technion, Haifa 32000, Israel\\
   E-mail: adi@physics.technion.ac.il}

\maketitle

\begin{abstract}
  We examine the implications of Modified Newtonian Dynamics (MOND) on
  the large scale structure in a Friedmann-Robertson-Walker universe.
  We employ a ``Jeans swindle'' to write a MOND-type relationship
  between the fluctuations in the density and the gravitational force,
  $\vg$. In linear Newtonian theory, $|\vg|$ decreases with time and
  eventually becomes $<\; g_{_0}$, the threshold below which MOND is
  dominant.  If the Newtonian initial density field has a power-law
  power-spectrum of index $n<-1$, then MOND domination proceeds from
  small to large scale.  At early times MOND tends to drive the density
  power-spectrum towards $k^{-1}$, independent of its shape in the
  Newtonian regime.  We use N-body simulations to solve the MOND
  equations of motion starting from initial conditions with a CDM
  power-spectrum.  MOND with the standard value $g_{_0}=
  10^{-8}{\mathrm cm\; s^{-2}}$, yields a high clustering amplitude
  that can match the observed galaxy distribution only with strong
  (anti-) biasing. A value of $g_{_0} \approx 10^{-9} {\mathrm cm\;
  s^{-2}}$, however, gives results similar to Newtonian dynamics and
  can be consistent with the observed large scale structure.

\end{abstract}
 
                           
\section{introduction}
Perhaps the best evidence for the existence of dark matter is the
flattening of rotation curves of many spiral galaxies.  Newton's
$g_{_N}\sim 1/({\rm distance})^2 $ law for the force of gravity
predicts that the observed distribution of baryonic matter should show
a fall off in the rotation curves, and hence the need for dark matter
to explain the flattening.  A universe predominantly filled with cold
dark matter (CDM) is consistent with observations of the cosmic
microwave background, and the galaxy population. Nevertheless, the
dark matter scenario does not readily explain the shape of rotation
curves of some galaxies in the inner regions (e.g., Moore 1994, Flores
\& Primack 1994, de Blok et al. 2001), the dark matter particle has not been discovered,
and we lack a direct experimental verification of Newton's law of
gravity at low accelerations.  So it seems prudent to examine
alternative scenarios explaining the flattening of rotation curves.
Modified Newtonian Dynamics (MOND) acting at low accelerations (e.g.,
Milgrom 1983, Bekenstein \& Milgrom 1984) can explain the flattening
of rotation curves without the need for dark matter.  Two main
versions of MOND exist. The first replaces Newton's law of gravity at
$g_{_N}<g_{_0} $\footnote{In MOND literature $a_{_0}$ is often used
instead of $g_{_0}$.}  by $g\sim \sqrt{g_{_0} g_{_N}}$ where
$g_{_0}\approx 1.2 \times 10^{-8}{\mathrm cm \; s^{-2}}$ is found to
give the best results in fitting the rotation curves. The second
modifies the law of inertia at low accelerations.  Here we adopt the
first version where the equations of motion can be derived from a
Lagrangian, admitting conservation of energy and momentum (Bekenstein
\& Milgrom 1984).  Because of the success of MOND at fitting the
rotation curves (e.g., Milgrom \& Braun 1988, Begeman et al. 1991,
Sanders 1996, de Blok \& McGaugh 1998, Sanders \& Verheijen 1998), it is worthwhile to push it further and confront it with
observations of the large scale structure in the Universe (Felten
1984, McGaugh 1999, Sanders 2001).  A problem here arises however.
MOND, when applied to a uniform background, predicts the collapse of
any finite region in the universe regardless of the mean density in
that region (e.g., Felten 1984, Sanders 1998).  To solve this problem
Sanders (2001) proposed a two-field Lagrangian based theory of MOND in
which the Friedmann-Robertson-Walker (FRW) background cosmology
remains intact in the absence of fluctuations. He argued that this
theory leads to large scale structure resembling Newtonian dynamics
with CDM-like initial conditions.  Here we take a simpler approach in
which we employ the Jeans swindle to write a MOND type relationship
between the fluctuations in the density and the gravitational force
field.  Although our procedure can be derived from a Lagrangian and is
equivalent to Sanders' two-field theory in the limit of small
coupling, we do not see any physical justification for it.
Nevertheless, the equations of motion have the desired properties and
we proceed to explore their consequences on the evolution of large
scale structure.

\section{Modification of the cosmological Newtonian equations of motion}
The background FRW cosmology is described by the scale factor $a(t)$
normalized to unity at the present, the Hubble function $H(t)=\dot a
/a$, and the total mean background density $\bar \rho_{tot} =\bar
\rho+\Lambda/(8\pi G)$, where $\bar \rho$ is the mean matter density
and $\Lambda$ is the cosmological constant.  Further, we define
$\Omega= \bar \rho/\rho_c$, where $\rho_c=3H^2/(8\pi G)$ is the
critical density. These cosmological quantities are related by
Einstein equations of general relativity.  Let $\vr $ and $\vx=\vr/a$
denote, respectively,  physical and comoving coordinates. The
fluctuations over the uniform background in the matter distributions
are described by the the comoving peculiar velocity $\vu=\dd \vx/\dd
t$ of a patch of matter, the density contrast $\delta(\vx) =\rho(\vx)
/\bar \rho-1$, where $\rho(\vx)$ is the local density, and the
fluctuations in the gravitational force field, $\vg$.
 Neglecting
 thermal effects, the Newtonian equations of motion governing the evolution
of the fluctuations are: the continuity equation
\begin{equation}
\frac{\pa \delta}{\pa t }+\vnabla_x \cdot (1+\delta) \vu =0 \; ,
\label{cont}
\end{equation}
the Euler equation of motion,
\begin{equation}
\frac{\dd \vu }{\dd t }+2H \vu =\vg /a\; ,
\label{euler}
\end{equation}
and the Poisson equation,
\begin{equation}
\frac{1}{a}\vnabla_x \cdot \vg =-4\pi G \bar \rho \delta=
-\frac{3}{2}\Omega H^2 \delta\; .
\label{poisson}
\end{equation}
In the linear regime these equations imply that $\vg|$
decreases with time (see 3.1).  
Once $|\vg| $ drops below $g_{_0}$, MOND takes over
the subsequent evolution of the fluctuations.  The
continuity equation (\ref{cont}) reflects mass conservations and holds
also in MOND. The MONdification (terminology by M. Milgrom) must be
done by either changing the law of inertia, i.e., the Euler equation
(\ref{euler}), or the relation between the density and the force field, i.e.,
the Poisson equation (\ref{poisson}).  Here we choose to 
maintain the Euler equation and 
replace the Poisson equation with
\begin{equation}
\frac{1}{a}\vnabla_x \cdot \left(\frac{ |\vg| }{g_{_0}} \vg\right)=
-\frac{3}{2}\Omega H^2 \delta \; .
\label{mondpoisson}
\end{equation}
In writing this equation we have assumed that MOND only affects the
fluctuations and leaves the background cosmology intact. This ``Jeans
Swindle'' (Binney \& Tremaine 1987)  is hard to justify on
physical grounds, even though it leads to equations of motion which can
be derived from a Lagrangian, admitting conservation of energy
and momentum (Sanders 2001, Bekenstein \& Milgrom 1984).  Although we
can take $g_{_0}$ to be a function of time here we assume it is a
constant and write it as
\begin{equation}
g_{_0}=f c H_{_0}=f R_H H_{_0}^2,
\end{equation}
where $f $ is a numerical fudge factor, $c$ is the speed of light,
$H_{_0}$ is the current value of  the Hubble function, and $R_H=c/H_{_0}$ is
the Hubble radius.  Fitting the rotation curves by MOND requires
$f\approx 1/6$ for $H_0=70 {\rm km \; s^{-1} \; Mpc^{-1}}$ (e.g. Sanders 1996), but we will see that $f$ must be
smaller for MOND to be in reasonable agreement with the observed large
scale structure.  Equation (\ref{mondpoisson}) determines $\vg$ only
up to a divergence free vector field $\vh$. So in a sense the theory
is incomplete. However, Bekenstein \& Milgrom (1984) have shown that
$\vh$ decays rapidly with increasing scale. So 
on large scales we neglect $\vh$ and assume
that $|\vg|\vg$ is irrotational so that it is fully determined by
(\ref{mondpoisson}). 
For conciseness we gather all the time dependent function in the Euler 
equation.
So let us define an irrotational vector field $ \vg_{_{N}}$
satisfying
\begin{equation}
\vnabla \cdot  \vg_{_N}=-\delta \label{sfreep} \;  .
\end{equation}  
Then, by (\ref{mondpoisson}), we have
\begin{equation}
\vg= 
\left(\frac{3}{2}a \Omega H^2g_{_0} \right)^{1/2}
\frac { { \vg}_{_N} }{
|{ \vg}_{_N}|^{1/2}}  \; .
\label{gntog}
\end{equation}
So  the Euler equation becomes
\begin{equation}
\frac{\dd \vu }{\dd t }+2H \vu =
\left(\frac{3}{2a} \Omega H^2g_{_0} \right)^{1/2}
\frac { { \vg}_{_N} }{
|{\vg}_{_N}|^{1/2}}  \; .
\label{eulerscaled}
\end{equation}
The Newtonian equivalent of this equation is
\begin{equation}
\frac{\dd \vu }{\dd t }+2H \vu =
\frac{3}{2}\Omega H^2  {\vg}_{_N}\; .
\end{equation}

\section{The limit of small density fluctuations
and early time MOND}

We treat now the MOND equations in the limit of small density
fluctuations. We focus on the evolution at early times when
$\Omega\approx 1$, restricting the treatment to a matter dominated
universe. Assuming that  the fluctuations are initially Newtonian,
we discuss the epoch at which they become MOND dominated, and the
shape of their power-spectrum in the MOND regime.

\begin{figure}
\centering
\mbox{\psfig{figure=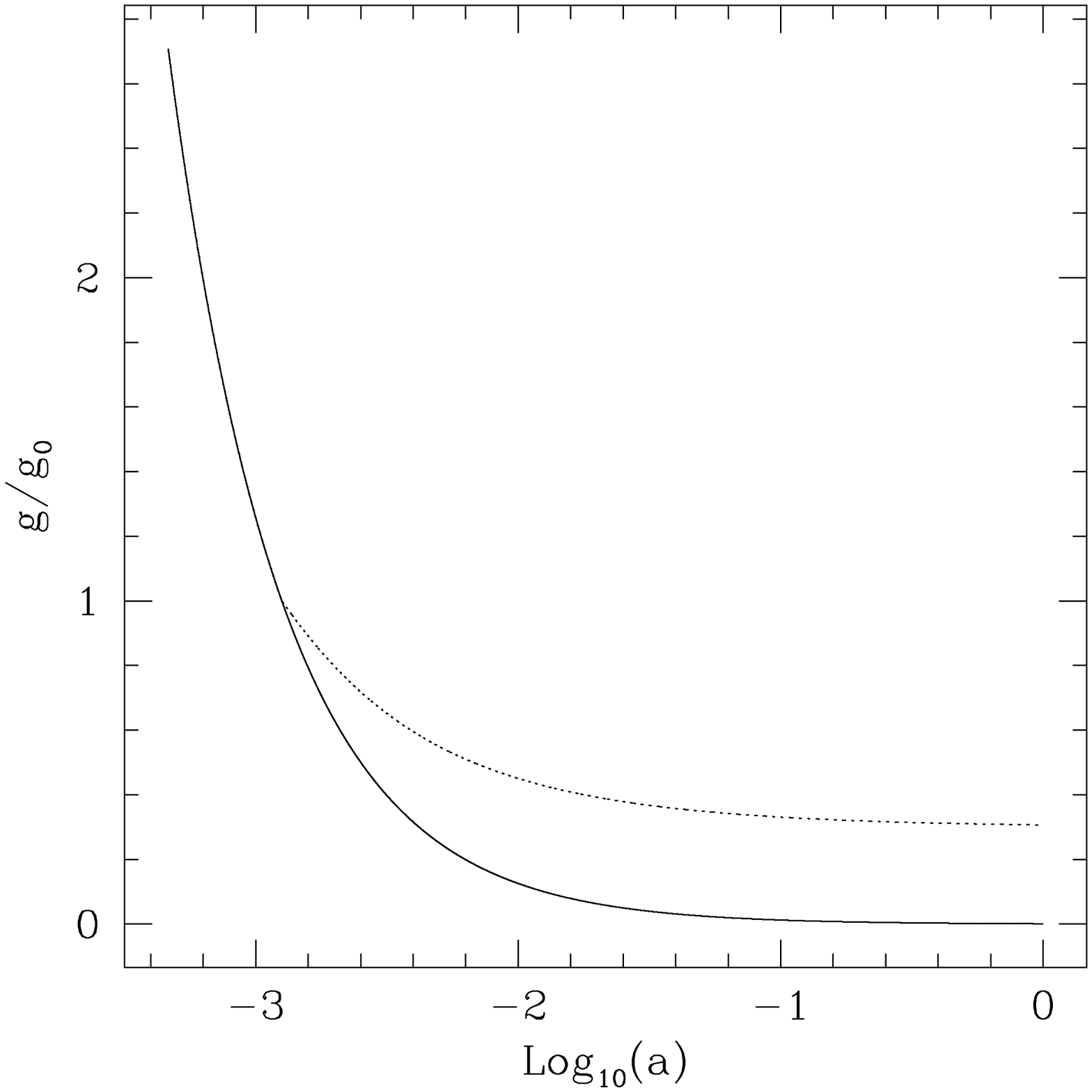,height=3.1in}}
\vspace{0.2cm}
\caption{The absolute value of $\vg/g_{_0}$ in the limit of small density 
fluctuations as a function of $a$ in an $\Omega=1$ matter dominated universe.  
The solid curve is the behavior according to  Newtonian dynamics.
The dotted  curve is obtained when MOND is activated  at $|\vg|<g_{_0}$.}
\label{glin}
\end{figure}

The modified Poisson equation (\ref{mondpoisson}) relating $\vg $ and
 $\delta$ is non-linear and cannot be linearized. However, we  can
 write the equations at early times when $\delta $ and$ |\vnabla \cdot
 \vu/H|$ are still $ \ll 1$. In this limit we neglect the term $\vnabla \cdot
 (\delta \vu)$ in the continuity equation (\ref{cont}) and replace the
 full derivative $\dd \vu /\dd t=\pa \vu /\dd t+\vu \cdot \vnabla \vu
 $ with $\pa \vu /\dd t$ in the Euler equation (\ref{eulerscaled}).
 The continuity equation then implies that $ \delta =-\vnabla \cdot
 \vu$ which in combination with equation (\ref{sfreep}) gives
\begin{equation}
\vu=\frac{\pa { \vg}_{_N}}{\pa t} \; .
\end{equation}
Substituting this result in the Euler equation (\ref{eulerscaled}) 
and neglecting $\vu \cdot \vnabla \vu$ in $\dd \vu /\dd t$ yield
\begin{equation}
\frac{\pa^2 { \vg}_{_N} }{\pa t^2 }+2H \frac{\pa  { \vg}_{_N} }{\pa t} =
\left(\frac{3}{2a} \Omega H^2 g_{_0}\right)^{1/2}
\frac{{ \vg}_{_N} }{|{ \vg}_{_N}|^{1/2}} \; .
\label{lineargn}
\end{equation}
This system of 3 first order differential equations determines
${ \vg}_{_N}$ locally at any point in space.  The full solution
depends on the values of ${ \vg}_{_N} $ and its time derivative at
some initial time, $t_i$.  We could not find analytic solutions to
this equation, but in an
$\Omega=1$ universe the asymptotic behavior  $ t\gg t_i$
can easily be obtained by substituting $\vg=\vb_{_0} t^\alpha$. This gives  
\begin{equation}
|{\vg}_{_N} |= \frac{27}{200}g_{_0} t_{_0}^{2} a^2(t) \; ,
\end{equation}
where $t_{_0}$ is the Hubble time. Using (\ref{gntog}), this  implies 
\begin{equation}
|{ \vg}|= \frac{3}{10} g_{_0} \; .
\end{equation}
So $|\vg|$ approaches a constant that is smaller than $g_{_0}$, i.e.,
fluctuations entering MOND never become Newtonian again.  
Integration of the MOND equations for the spherical collapse model (not shown
here),  and
the N-body simulations presented in section 4  show that 
$|\vg|<g_{_0}$ also in the moderately non-linear regime. 
In Fig. \ref{glin} we plot $|\vg|/g_{_0}$ as a function
of $a$, in the limit of small density fluctuations in an $\Omega=1$
universe.  The solid line follows the growing mode of linear Newtonian
theory.  The dotted line shows $|\vg|$ in the MOND regime and is
obtained by numerical integration of (\ref{lineargn}) for $|\vg|<g_{_0}$.
This numerical solution confirms our analytic result that $|\vg|$
approaches $(3/10)g_{_0}$.  This asymptotic behavior has interesting
implications on the topology of the density field.  It implies that
one-dimensional pancake-like perturbations the density contrast
$\delta=-\vnabla \cdot {\vg}_{_N} $ vanishes in all space except
at the surfaces of $\vg=0$, where it can be described by a Dirac-delta
function. In a generic three-dimensional perturbation, surfaces of
$\vg=0$ are still singular but the density contrast in the rest of
space is non-vanishing and grows with time like $\delta\sim
t^{4/3}\sim a^2$ as compared to $\delta\sim a$ for the growing mode in
linear Newtonian theory (e.g. Peebles 1980).  We will use this result
in 3.2 to estimate the shape of the power-spectrum in early time MOND.

\subsection{Epoch of MOND domination}

The fluctuations in the Newtonian gravitational force field over a
spherical region of physical radius $r=a x$ and density contrast
$\delta$ are $g_{_N}\sim G\delta \bar M/(a x)^2$, where $\bar M\sim \bar
\rho (a x)^3$.  Since $\bar \rho \sim 1/a^3$, $g_{_N}$ decreases with
time if $\delta \sim t^\alpha$ with $\alpha <2$.  In linear Newtonian
theory, $\alpha \le 1$ and so at early enough times $g_{_N}>g_{_0}$ and the
fluctuations would be in the Newtonian regime.  At later times $g_{_N}$
drops below $g_{_0}$ and the fluctuations enter the MOND regime.  Here we
determine the redshift at which fluctuations  on a comoving
scale $x$ become MOND dominated.  Assume that the transition epoch between Newtonian
dynamics and MOND occurs early enough so that $\Omega\approx 1$, an
assumption that will be justified below.  The initial density field in
the Newtonian regime is gaussian and characterized by the
power-spectrum, $<|\delta_k|^2>$, where $\delta_k$ is the Fourier
transform of the density as a function of the wavenumber, $k$. If the
power-spectrum is a power-law of index $-3<n<1$, then the $rms$ value
of the density field smoothed on a comoving scale $x$ at redshift $z$ is
approximately
\begin{equation}
\sigma(x,z)=a(z) \left(\frac{x}{x_{_0}}\right)^{-\frac{n+3}{2}}\; ,
\label{sigma}
\end{equation}  
where the redshift dependence through $a(z)=1/(1+z)$ follows from
linear Newtonian theory if $\Omega=1$, and the scale $x_{_0}$ sets the
amplitude of the fluctuations.  According to either (\ref{poisson}) or
$g_{_N}\sim G \delta \bar M/{r^2}$, the $rms$ of $g_{_N}$ on a scale $x$ is
\begin{equation}
\sigma_g\sim \frac{3}{2} H^2 a x \sigma \; .
\label{sigmag}
\end{equation} 
The redshift, $z_{_{\rm mond}}$, at which 
$g_{_N}=g_{_0}=fR_H H_{_0}^2$ is then
\begin{equation}
z_{_{\rm mond}}
=\frac{2}{3}f \frac{R_H}{x_{_0}}\left(\frac{x}{x_{_0}}\right)^{\frac{n+1}{2}}
-1 \; .
\label{mondtrans}
\end{equation}
So if $n<-1$ then fluctuations on small scales enter the MOND regime
before large scale fluctuations.  For $n>-1$ the interpretation that
large scales become MOND dominated earlier is somewhat vague. It is
hard to visualize a large scale fluctuation in the MOND regime while
the superimposed small scale fluctuations are still Newtonian. A more
appropriate interpretation is that for $n>-1$ the volume of space
containing fluctuations in the MOND regime is mainly made up of large
regions separated by small scale fluctuations that are still
Newtonian.  This interpretation is sustained by the following
argument.  Since in the Newtonian regime $\vg$ is gaussian, the
probability that a spherical region of radius $x$ becomes  MOND dominated 
at time $t$ is,
\begin{equation}
P\propto \int_{_0}^{g_{_0}/\sigma_g}\exp\left(-y^2/2\right)y^2\dd y  \; .
\end{equation}
According to (\ref{sigmag}), $\sigma_g\propto x^{-(n+1)/2}$
so if $n>-1$ then $P$ increases with scale, meaning that the fraction
of space already in MOND is made up of large regions. If $n<-1$ then
$P$ decreases with scale and MOND penetrates small regions before it
dominates the whole space.

Fluctuations enter the MOND regime at a relatively early time.  Taking
$x_{_0} \sim 10\mpc $ and $ R_H\sim 3000 \mpc$ then, according to
(\ref{mondtrans}), the transition into MOND occurs at $z> f 200$, on
scales $\gtsim 10 \mpc$ if $n>-1$, and $\ltsim 10 \mpc$ if $n<-1$.  In
a CDM-like power-spectrum the effective power-index on scales $\gtsim
20\mpc$ is $n\gtsim-1$ and changes to $-3 <n\ltsim -1$ on smaller
scales.  For this type of power-spectrum the transition into MOND of
all scales occurs at an early time when the density fluctuations are
still small and $\Omega\approx 1$. This considerably simplifies our
task of estimating  the density power-spectrum  in the MOND
regime in the next subsection.

\subsection{power-spectrum in the MOND regime at early times}
MOND gives accelerations $\propto \sqrt{g_{_0} g_{_N}}> g_{_N} $, leading to
faster growth rates than Newtonian gravity.  So we expect a break in
the shape of the power-spectrum at the scale just entering MOND.
Assuming a power-law power-spectrum, $<|\delta_k|^2>\sim k^n$, in the
Newtonian regime we estimate the power-index of the density
fluctuations after they become MOND dominated.  Consider the $rms$
value, $\tilde \sigma(x,a)$, of the density field over any scale $x$.
Suppose that at redshift $z_i\gg 1$ fluctuations on all scales were
Newtonian so that $\tilde \sigma=\sigma(x,z_i)$, where $\sigma$ is
given by (\ref{sigma}). Let $\xm(z)$ be the scale just entering the
MOND regime at redshift $z<z_i$.  For simplicity we consider the case
$n<-1$, so at $x> \xm$ the evolution is Newtonian and $\tilde \sigma
=\sigma$.  A fluctuation on a scale $x< \xm$ entered MOND at redshift
$z_x>z$.  As we have seen in the previous subsection fluctuations on a
wide range of scales become MOND dominated at an early enough time so
that $\Omega\approx 1$. In an $\Omega\approx 1$ universe, the growth
factors in the Newtonian and MOND regimes are $\delta\sim a$ and
$\delta\sim a^2$, respectively.  So, from $z_i $ until $z_x$ the
fluctuation grew by a factor of $(a_x/a_i)$, while from $z_x$ until
$z$ it grew by $(a/a_x)^2$, where $a_i=a(z_i)$, $a_x=a(z_x)$, and
$a=a(z)$. So for $x<\xm$,
\begin{equation}
\tilde \sigma(x,a)=\left(\frac{a}{a_x}\right)^2\left( \frac{a_x}{a_i}\right)
\sigma(x,a_i) \; .
\end{equation}
According to  (\ref{mondtrans}) 
\begin{equation}
\frac{a}{a_x}= \left(\frac{\xm}{x}\right)^{-\frac{n+1}{2}} 
\end{equation}
and
\begin{equation}
\tilde \sigma(x,a)=a \left(  \frac{\xm}{x_{_0}}\right)^{- \frac{n+1}{2}}
\left(\frac{x}{x_{_0}}\right)^{-1} \; ,
\end{equation}
so  the power-index in the MOND regime is $n_{_{\rm
    mond}}=-1$. A similar argument for $ n>-1$ yields the same result.
Therefore the power-index in the MOND regime is independent of the value of
$n$.  A general power-spectrum can be approximated as a power-law over
any limited range of scales where our result applies.  Therefore, MOND
drives the power-spectrum towards $k^{-1}$ independent of its shape in
the Newtonian regime. 

The treatment here ignored mode-coupling 
and  assumed
that all fluctuations became MOND dominated when  $\Omega$ was still
very close to unity. 
Although the simulations  presented in
the next section do not show strong mode-coupling for low values of $g_{_0}$,
 the shape of the power-spectrum derived here 
should  serve as a general indication only.

\section{Results from N-body simulations}

In this
section we study the general evolution of fluctuations under MOND by
means of N-body simulations. This is done by adapting a particle-mesh
(PM) N-body code originally written by E.  Bertschinger to solve the
Newtonian equations (Bertschinger \& Gelb 1991).  At each time step,
the code readily provides the irrotational component of ${\vg}_{_N}$
from the density field as estimated on a grid from the particle
distribution.  We adapt the code by adding a routine to compute the
gravitational field $\vg$ from ${\vg}_{_N}$ using the relation
(\ref{gntog}). We also incorporate the time dependent functions in the
MOND equations in a leapfrog time integration scheme to move the
particles to the next time step.

\begin{table}
\caption{Details of the six simulations.
 Each simulation contained
$64^3$ particles in a cubic box of $128 {\mathrm h}^{-1} {\mathrm
Mpc}$ on the side. The $rms$ value, $\sigma_8$, of density fluctuations in
spheres of radius $8\mpc$ is measured from the output of each
simulation at the final time.}
\begin{tabular}{|l|c|c|c|c|} 
simulation& gravity & $g_{_0}/(1.2\times 10^{-8}{\mathrm cm /s}^2)$ &$\Omega_{0}$&$\sigma_8$\\\hline
I & MOND& 1& 1 &3.9\\\hline
II & MOND& 1& 0.03 &4.83\\\hline
III & MOND& 1& 1 & 1.16 \\
 & & ($10^{-6}\delta_{\mathrm initial}$)& \\\hline
IV & MOND& 1/12& 1 & 1.08\\\hline
V & MOND& 1/12& 0.03 &1.01\\\hline
VI & Newton& 0& 1& 0.86 \\\hline
\end{tabular}
\label{tab1}
\end{table}
Five MOND simulations with different values of $g_{_0}$ and
$\Omega_{0}\equiv \Omega(a=1)$ were run. For comparison, a simulation
with Newtonian gravity was also run.  The simulations started with
gaussian initial conditions generated from the CDM power-spectrum with
$H_{_0}=50\mpc$ and $\Omega_{0}=1$.  Each simulation contained $64^3$
particles in a cubic box of $128\mpc$ on the side. The CDM power
spectrum on the large scales probed by the box is close to $k^{-1}$ as
motivated by our result for the density power-spectrum in the MOND
regime at early times.  The parameters of each simulation are written
in Table 1.  Except simulation III, all simulations started with the
same initial conditions which were normalized so that the density
field in Newtonian simulation at the final time had an $rms$ value of
$\sigma_8=0.86$ in spheres of radius $8\mpc$.  In simulation III, the
amplitude of the initial conditions was reduced by a factor of $10^6$.
Simulations with $\Omega_0=1$ and $0.03$ started at $a=0.049$ and $0.01$,
respectively. At these two time the linear Newtonian density growth
factor, $D$, was the same for the two values of $\Omega_0$.  For the
high $g_{_0}$, this choice of the starting time of the simulations
corresponds roughly to the epoch at which fluctuations on all scales
become MOND dominated according to our estimate in section 3.2. The
particle distribution at this time was obtained using the Zel'dovich
approximation.  We are interested in comparing the net dynamical
effect of decreasing $g_{_0}$ and so we start MOND simulations with
low and high $g_{_0}$ at the same value of $D$. 

\begin{figure}
\hspace{2.5cm}
%  \mbox{\psfig{height=5.in,width=0.6\textwidth}}
  \vspace{1.1cm}
\caption{The particle distribution in a slice of thickness $2.56\mpc$
  through the simulations at the final time.  The panels in a
  clockwise order correspond to simulations I---VI listed in the table
  1. {\bf See fig2.gif}}
\label{slice0}
\end{figure}

\begin{figure}
\centering
\mbox{\psfig{figure=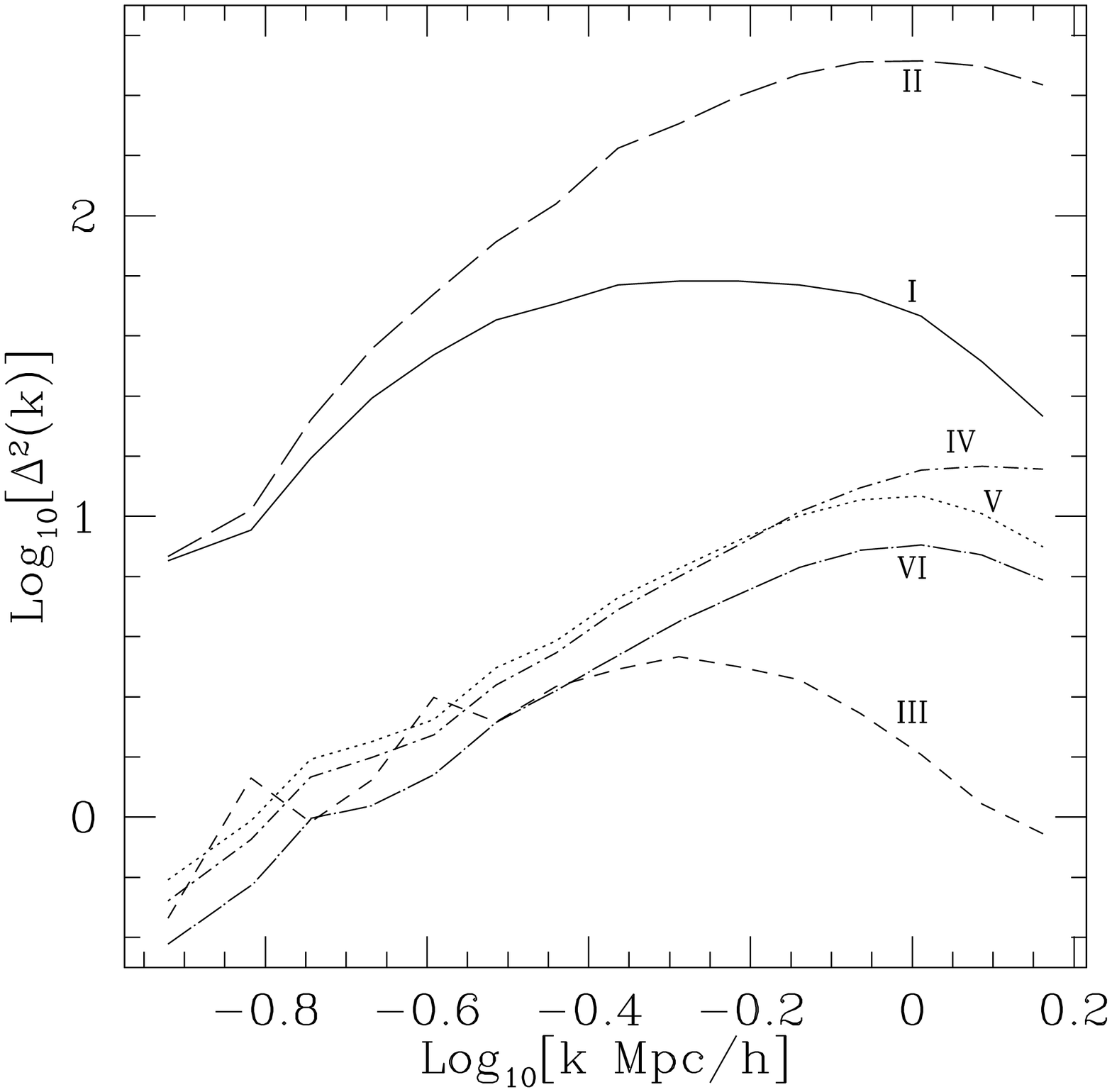,height=3.1in}}
\vspace{0.1cm}
\caption{The variance $\Delta^2(k)=k^3<|\delta_k|^2>/(2\pi^2)$ of 
  density fields per $\ln(k)$ from the simulations at the final time.  The
  solid, long-dashed, short-dashed, dotted, dot-short-dashed, and
  dot-long-dashed and lines correspond respectively to simulations
  I--VI. }
\label{pk}
\end{figure}

Fig. \ref{slice0} show the particle distribution in a slice in the
simulations at $a=1$.  This figure demonstrates that MOND with the
standard value of $g_{_0}\approx 1.2 \times 10^{-8}{\mathrm cm
s^{-2}}$ (bottom and middle panels to the left) yields high clustering
at the present time, for both $\Omega_{0}=1$ and $0.03$. Lowering the
amplitude of the initial fluctuations by a factor of $10^6$ reduces
the clustering at the present time but severely washes out small scale
fluctuations (top panel to the left).  For both values of
$\Omega_{0}$, MOND with a smaller value for $g_{_0}$ produces particle
distribution (top and middle panels to the right) similar to Newtonian
dynamics (bottom panel to the right). The values of $\sigma_8$
measured from simulations at the final time are listed in Table 1.
Fig. \ref{pk} shows $\Delta^2 (k)=V k^3 <|\delta_k|^2>/2\pi^2$ where
$V$ is the volume of the simulation box (Peacock 1999). This quantity
gives the contribution to the variance per $\ln(k)$.  The MOND curves
with low $g_{_0}$ (IV and V) follow the shape of the Newtonian curve
(VI).

\begin{figure}
\centering
%\mbox{\psfig{figure=gv.ps,height=3.1in}}
\vspace{0.1cm}
\caption{The peculiar velocity {\it vs.}  the ``Newtonian'' gravity
field $\vg_{_N}$ obtained from the density contrast using
(\ref{sfreep}).  The panels show fields from simulations II
($\Omega_{0}=0.03$, high $g_{_0}$), IV ($\Omega_{0}=1$, low $g_{_0}$),
V ($\Omega_{0}=0.03$, low $g_{_0}$), and VI (Newtonian), as indicated
in the figure. The fields were smoothed with a Gaussian window of
radius $8\mpc$. The diagonal line in each panel is plotted to guide
the eye. {\bf See fig4.gif}}
\label{gv}
\end{figure}

A common method for estimating the cosmological density parameter,
$\Omega$, is the comparison between the measured peculiar velocity
field and the ``Newtonian'' gravity field, $\vg_{_N}$, inferred
observed distribution of galaxies using (\ref{sfreep}) (see Strauss \&
Willick 1995 and references therein).  Putting aside the issue of
redshift distortions, linear Newtonian theory gives $f(\Omega)\vu=H
\vg_{_N}$, where $f(\Omega)\approx \Omega^{0.6}$ (e.g., Peebles 1980).
This comparison has been done using several data sets with the
conclusion that the observed velocity and gravity fields are tightly
correlated. To test whether MOND can account for the correlation seen
in the observations, we have performed a comparison between the
velocity and ``Newtonian'' gravity field, $\vg_{_N}$, computed from
the simulations at the final time.  The results are shown in Fig.
\ref{gv}. To avoid strong non-linearities, the fields were smoothed
with a gaussian window of radius $8\mpc$.  The $rms$ values of the one
dimensional $\vu$ and $H \vg_{_N}$ shown in the figure are
$(\sigma_u,\sigma_g)=(1176,336)$, (450,302), (223,336), and (209,226)
in ${\rm km \; s^{-1} }$, respectively, for the simulations II, IV, V,
and VI.  Simulation II is highly non-linear even on scales larger than
$8\mpc$ and the relative scatter between the fields is large. In the
simulations with small $g_{_0}$, the correlation between the fields is
tight for both $\Omega_0=1$ (simulation IV) and $\Omega_0=0.03$ (V).
The scatter between the fields is larger than in the Newtonian
simulation (VI), but it is negligible compared to observational
uncertainties.  The slope of the regression of $u_x$ on $H g_{_{N,x}}$
is higher in simulation $IV$ than in the Newtonian simulation,
$VI$. These two simulations have $\Omega_0=1$ but as is the case with
rotational speeds of galaxies, MOND also accounts for larger
velocities on large scales.  These results indicate that MOND can be
consistent with this type of analysis. In MOND, the slope of the
regression of $\vu$ on $H \vg_{_N}$ is also a decreasing function of
$\Omega$, but is not proportional to $f(\Omega)$. For example for the
MOND simulation V with low $g_{_0}$ and $\Omega_0=0.03$, the slope is
0.55, instead of $0.25 $ in linear Newtonian theory.  Note that the
relation (\ref{gntog}) between $\vg$ and $\vg_{_N}$ implies that
$|\vg|\ll g{_{0}}$ in all MOND simulations, sustaining our conjecture
that large scale fluctuations never leave the MOND regime.
We have also computed the probability distribution function (PDF) of
density field smoothed on $8\mpc$. The shape of the PDF in the
simulations with small $g_{_0}$ is very close to that in the Newtonian
simulation.

The simulations shown so far were run assuming all particles enter the
MOND regime at the same time. This captures the net dynamical effects
of varying $g_{_0}$. However, a realistic MOND simulation should start
at an early enough time when most particles are still Newtonian.  A
particle moves accroding to Newtonian dynamics, until the its peculiar
gravitational (in absolute value), $g$, drops below $g_{_0}$. The
subsequent motion of the particle is governed by MOND as long as
$g<g_0$.  We have run a suite of simulations with this recipe for
modeling the transition into MOND. Of course the transition into MOND
could be made a little smoother but in the lack of an exact model for
how the transition occurs we choose to proceed assuming sudden
transition.  The simulations we have run correspond to the CDM
power-spectrum used in the simulations shown in Fig. \ref{slice0}, and
two scale-free power-spectra with power-indices, $n=-1$ and $n=1$.
All simulations were run for $\Omega=0.03$ with the standard and low
value of $g_0$.  The initial conditions were normalized similarly to
the Newtonian simulation in table 1.  Slices of the particle
distribution in these simulations are shown in Figs. \ref{slicef0} and
\ref{slicef01}, for $a=1$ and $a=0.1$, respectively. The particle
distribution of the CDM simulation in Fig. \ref{slicef0} (bottom
panels) is a little more evolved than in Fig.  \ref{slice0}. Despite
this, the particle distributions in the two simulations are very
similar.  This indicates that MOND becomes dominated at a similar
epoch for all particles and that our estimate of this epoch used in
the simulations shown in Fig.  \ref{slice0} is reliable. 
\begin{figure}
\hspace{2.5cm}
%  \mbox{\psfig{figure=xyf20.ps,height=5.in,width=0.6\textwidth}}
  \vspace{1.1cm}
\caption{The particle distribution in a slice of thickness $2.56\mpc$
through the simulations at the final time, $a=1$.  These simulations
have $\Omega_0=0.03$ and 
started early enough when most particle were Newtonian.  A particle
became MOND dominated once its peculiar acceleration dropped below
$g_0$. The panels on the left are for simulations run with standard
$g_0$ while the panels on the right are with low $g_0$ (1/12
standard). Bottom, middle, and top rows correspond to power-spectra of
CDM, $n=-1$, and $n=+1$, respectively. {\bf See fig5.gif}}
\label{slicef0}
\end{figure}
\begin{figure}
\hspace{2.5cm}
%  \mbox{\psfig{figure=xyf11.ps,height=5.in,width=0.6\textwidth}}
  \vspace{1.1cm}
\caption{The same as the previous figure (\ref{slicef0}), but 
for simulations at $a=0.1$. {\bf See fig6.gif}}
\label{slicef01}
\end{figure}

\section{Summary and discussion}
Several arguments against MOND have been presented in the literature (
e.g., Felten 1984,  van den Bosch et al. 2000, Mortlock \& Turner
2001, Aguirre et al. 2001, Scott et al. 2001).  The reader may decide
for her/himself which of these arguments are compelling.  Here we
focus on examining the general properties of large scale structure
under MOND and put aside the arguments presented in the literature
against MOND.  The value of $g_{_0}$ necessary to fit the rotation
curves of galaxies produces far too much clustering and can be
consistent with observations only with strong anti-bias.  Decreasing
$g_{_0}$ has two effects: 1) it pushes the epoch of MOND domination to
later time, and 2) it reduces the typical accelerations produced by
MOND forces.  The simulations presented in this paper show that MOND
with smaller $g_{_0}$ leads to large scale structure similar to
Newtonian dynamics. This is encouraging for MOND since the standard
paradigm based on Newtonian gravity and dark matter is successful at
explaining many observations of the large scale structure and the
galaxy population (e.g., Kauffman, Nusser \& Steinmetz 1997, Benson et
al. 2000, Diaferio et al. 2001) . Still one has to adopt a
generalization of MOND, allowing for, or at least mimicking, a scale
dependent $g_{_0}$.  Sanders' (2001) two-field theory for example can
serve as a general framework for that purpose.  In this theory the
rapid growth of fluctuations in MOND is tamed by the coupling between
the two fields.  The amplitude of the initial conditions used in the
simulations roughly matches the normalization implied by measurements
of the cosmic microwave background anisotropies (Hu W., Sugiyama \&
Silk 1997).  We have seen that by reducing the amplitude of the
initial conditions by a large factor MOND with the standard value of
$g_{_0}$ produces the correct $\sigma_8$, but causes a severe washing
out of structure on scale of 10s of $\mpc$. Therefore it seems
necessary to decrease $g_{_0}$ to match the large scale structure.

Our goal here was to present general criteria for the consistency of
MOND with large scale structure. Further study of MOND within a
realsitic recipe for galaxy formation to clarify the role of (anti-)
biasing remains to be done.

\section{Acknowledgments} 
I thank Ed Bertschinger for allowing the use of his PM code, Simon White and 
Stacey McGaugh for useful comment.  This
research was supported by the Technion V.P.R Fund- 
Henri Gutwirth Promotion of Research Fund, and the German Israeli Foundation
for Scientific Research and Development.  
\protect

\bigskip
\begin{thebibliography}{}
\bibitem{} Aguirre A., Schaye J., Quataert E., 2001, accepted, ApJ, 
astro-ph/0105184
\bibitem{} Begeman K.G., Broeils A.H.,  Sanders R.H. 1991, MNRAS, 249, 523
\bibitem{} Bekenstein J., Milgrom M., 1984, ApJ,  286, 7
\bibitem{} Benson A.J., Cole S., Frenk C.S., 
Baugh C.M., Lacey C.G., 2000, MNRAS, 311, 793
\bibitem{} Bertschinger E., Gelb J.M, 1991, Computers in Physics, 5, 164
\bibitem{} Binney J.J., Tremaine S., 1987, {\it Galactic Dynamics}, Princeton 
University Press
\bibitem{} de Blok W.J.G., McGaugh S.S.,  1998, ApJ, 508, 132
\bibitem{} de Blok W.J.G., McGaugh S.S., Bosma A., Rubin V.C., 2001, 
ApJL, 552, 23
\bibitem{} Diaferio A.,  Kauffman G., Balogh M.L., White S.D.M., Schade D., 
Ellingson E., 2001, MNRAS, 323, 999
\bibitem{} Felten J.E., 1984, ApJ, 286, 3
\bibitem{} Flores R.A. Primack J.P., 1994, ApJ, 427,1 
\bibitem{} Hu W., Sugiyama N., Silk J., 1997, Nature, 386, 37
\bibitem{} Kauffmann G., Nusser A., Steinmetz M., 1997, MNRAS, 286, 795
\bibitem{} Milgrom M., 1983, 270, 365
\bibitem{} Milgrom M., Braun E., 1988, ApJ, 334, 130
\bibitem{} McGaugh S.S, 1999, ApJ, 523, L99
\bibitem{} Moore B.,1994, Nature, 370, 629
\bibitem{} Mortlock D.J., Turner E.L., 2001, MNRAS, 327, 557
\bibitem{} Peacock J.A, 1999, {\it Cosmological Physics}, Cambridge University 
Press
\bibitem{} Peebles J.P.E, 1980 {\it Large Scale Structure}, Princeton 
University Press

\bibitem{} Sanders R.H., 1996, ApJ, 473, 117
\bibitem{} Sanders R.H., 1998, MNRAS, 296, 1009
\bibitem{} Sanders R.H., Verheijen M.A.W.,  1998, ApJ, 503, 97
\bibitem{} Sanders R.H., 2001, astro-ph/0011439
\bibitem{} Scott D., White M., Cohn J., Pierpaoli E., 2001, astro-ph/014435
\bibitem{} Strauss M.A., Willick J.A, 1995, Physics Reports, 261,271

\bibitem{} van den Bosch F.C., Dalcanton J.J., 2000, ApJ, 534, 146
\end{thebibliography}
\end{document}